\documentclass[11pt]{article}
\usepackage{winter,epsfig}
\usepackage{rotating}

\def\Journal#1#2#3#4{{#1} {\bf #2}, #3 (#4)}

\def\PLB{{\em Phys. Lett.}  B}
\def\PRL{\em Phys. Rev. Lett.}
\def\PRD{{\em Phys. Rev.} D}
\def\ZPC{{\em Z. Phys.} C}
\def\EPJC{{\em Eur. Phys. J.} C}
\def\CPC{\em Comput. Phys. Commun.}
\newcommand {\epem} {\mathrm{e^+e^-}}
\newcommand {\Z}  {\mathrm Z}

\newcommand {\x}    {{\mathrm X_0}}
\newcommand {\deltaZ} {\delta_{\mathrm{Z}}}
\newcommand {\LEPone} {LEP\,1}
\newcommand {\LEPtwo} {LEP\,2}
\newcommand{\phz} {\phantom{0}}

\begin{document}
\vspace*{4cm}
\title{RUNNING OF THE QED COUPLING IN SMALL-ANGLE BHABHA SCATTERING AT LEP}

\author{ G. ABBIENDI }

\address
{INFN and Dipartimento di Fisica dell' Universit\`a di Bologna, 
Viale C.~Berti-Pichat 6/2, \\ 
40127 Bologna, Italy}

\maketitle\abstracts{
Using the OPAL detector at LEP, the running of the effective QED
coupling $\alpha(t)$ is measured for space-like momentum transfer,
$2 \leq -t \leq 6$~GeV$^2$, 
from the angular distribution of small-angle Bhabha scattering. 
This is currently the most significant direct observation of the 
running of the QED coupling in a single experiment
and the first clear evidence
of the hadronic contribution to the running in the space-like region.
Our result is in good agreement with standard evaluations of $\alpha(t)$,
based on data in the time-like region.
}

\section{Introduction}
The effective QED coupling $\alpha(t)$ is an essential ingredient for many
precision physics predictions. 
It contributes one of the dominant uncertainties
in the electroweak fits constraining the Higgs mass. %\cite{lepew04}.
The effective QED coupling is generally expressed as:
\begin{equation}
\alpha(t) = \frac{\alpha_0}{1-\Delta\alpha(t)} 
\end{equation}
where $\alpha_0 = \alpha(t=0) \simeq 1/137$ 
is the fine structure constant, $t$ is the momentum transfer
squared of the exchanged photon and $\Delta\alpha$ is the vacuum polarization
contribution. 
Whereas the leptonic contributions to $\Delta\alpha$ are calculable 
to very high accuracy, the hadronic ones have to be evaluated 
by using a dispersion integral over the measured cross section of 
$\epem \rightarrow \mathrm{hadrons}$ at low energies, 
plus perturbative QCD \cite{ej95,bp2001}.
There are also many evaluations which
are more theory-driven, extending the application of perturbative QCD down to
$\sim 2$~GeV (see for example the reference \cite{detroconiz}).
An alternative approach \cite{jeger} uses perturbative QCD in the negative $t$ 
({\em space-like}) region.

There have been only a few direct observations of the running 
of the QED coupling \cite{topaz,opalaem,venus,l3}.
Here we present a new result from the OPAL collaboration. 
A full description can be found in the OPAL paper \cite{alphat}.
The running of $\alpha$ is measured in the space-like region,
by studying the angular dependence of small-angle Bhabha scattering,
$\epem \rightarrow \epem$, at LEP.
Small-angle Bhabha scattering appears to be an ideal process
for a direct measurement of the running of $\alpha(t)$ in a
single experiment, as it is an almost pure QED process, 
strongly dominated by $t$-channel photon exchange.  
Moreover the data sample has large statistics and excellent purity.
The Bhabha differential cross section can be written in the following form
for small scattering angle:
\begin{equation}
\frac{\mathrm{d}\sigma}{\mathrm{d}t} = 
\frac{\mathrm{d}\sigma^{(0)}}{\mathrm{d}t}
{\left( \frac{\alpha(t)}{\alpha_0} \right) }^2 (1+\epsilon)~(1+\delta_{\gamma})
+ \deltaZ
\label{eq:xsec}
\end{equation}
where ${\mathrm{d}\sigma^{(0)}}/{\mathrm{d}t} = {4 \pi \alpha_0^2}/{t^2}$
is the Born term for the $t$-channel diagram, $ \epsilon $ represents the
radiative corrections to the Born cross section, while
$\delta_{\gamma}$ and $\deltaZ$ are the interference contributions
with $s$-channel photon and $\Z$ exchange respectively.
$\delta_{\gamma}$ and $\deltaZ$ are much smaller than $\epsilon$
and the vacuum polarization. 
Therefore, with a precise knowledge of
the radiative corrections ($\epsilon$ term) one can determine the effective
coupling $\alpha(t)$ by measuring the differential cross section.
This method has also been advocated in a recent paper
\cite{arbuzov}.

\section{Detector and event selection}
We use OPAL data collected in 1993-95 
at energies close to the $\Z$ resonance peak.
In particular this analysis is based on 
the OPAL SiW luminometer \cite{lumipap}.
The SiW consisted of two cylindrical calorimeters 
encircling the beam pipe at a distance
$z \simeq \pm 2.5$~m from the interaction point. 
Each calorimeter was a stack of 19 silicon layers interleaved with 18
tungsten plates, with a sensitive depth of 14~cm, representing 22 radiation
lengths ($\x$).
The sensitive area fully covered radii between 6.2
and 14.2~cm from the beam axis, 
corresponding to scattering angles between 25 and 58 mrad. 
Each detector layer was segmented with $R$-$\phi$ geometry in a 
$32 \times 32$ pad array. The pad size was 2.5~mm radially and 11.25 degrees in
azimuth. In total the whole luminometer had 38,912 readout channels
corresponding to the individual silicon pads. 
Particles coming from the interaction point had to traverse the material
constituting the beam pipe and its support structures 
as well as detector cables before reaching the face of the SiW calorimeters. 
This preshowering material was minimum near the inner angular limit, 
about $0.25\,\x$, while in the middle of the acceptance 
it increased to about $2\,\x$.
When \LEPtwo\ data-taking started in 1996 the detector configuration changed,
with the installation of tungsten shields designed to protect the inner tracking
detectors from synchrotron radiation. 
This reduced the useful acceptance of the detector at the
lower angular limit. Therefore we limited this analysis to the 
\LEPone\ data.

The event selection is similar to the one used for luminosity measurements
\cite{lumipap}. 
The selected sample is strongly dominated by two-cluster configurations,
with almost full energy back-to-back $\mathrm e^+$ and $\mathrm e^-$ 
incident on the two calorimeters. 
At leading order the momentum transfer squared $t$ 
is simply related to the scattering angle $\theta$, which is measured from 
the radial position $R$ of
the scattered $\mathrm e^+$ and $\mathrm e^-$ at reference planes 
located within the SiW luminometers:
\begin{equation}
t = - s \, \frac{1- \cos\theta}{2} \approx - \frac{s \,\theta^2}{4} \,;
{\phantom{000000} } \tan \theta = R / z \,.
\label{eq:t}
\end{equation}
At the center-of-mass energy $\sqrt{s} \approx 91$~GeV our angular acceptance
corresponds to $2 \leq -t \leq 6$~GeV$^2$.

\begin{figure}
\begin{center}
\begin{tabular}{cc}
\epsfxsize=0.46\textwidth\epsfbox[0 0 567 680]{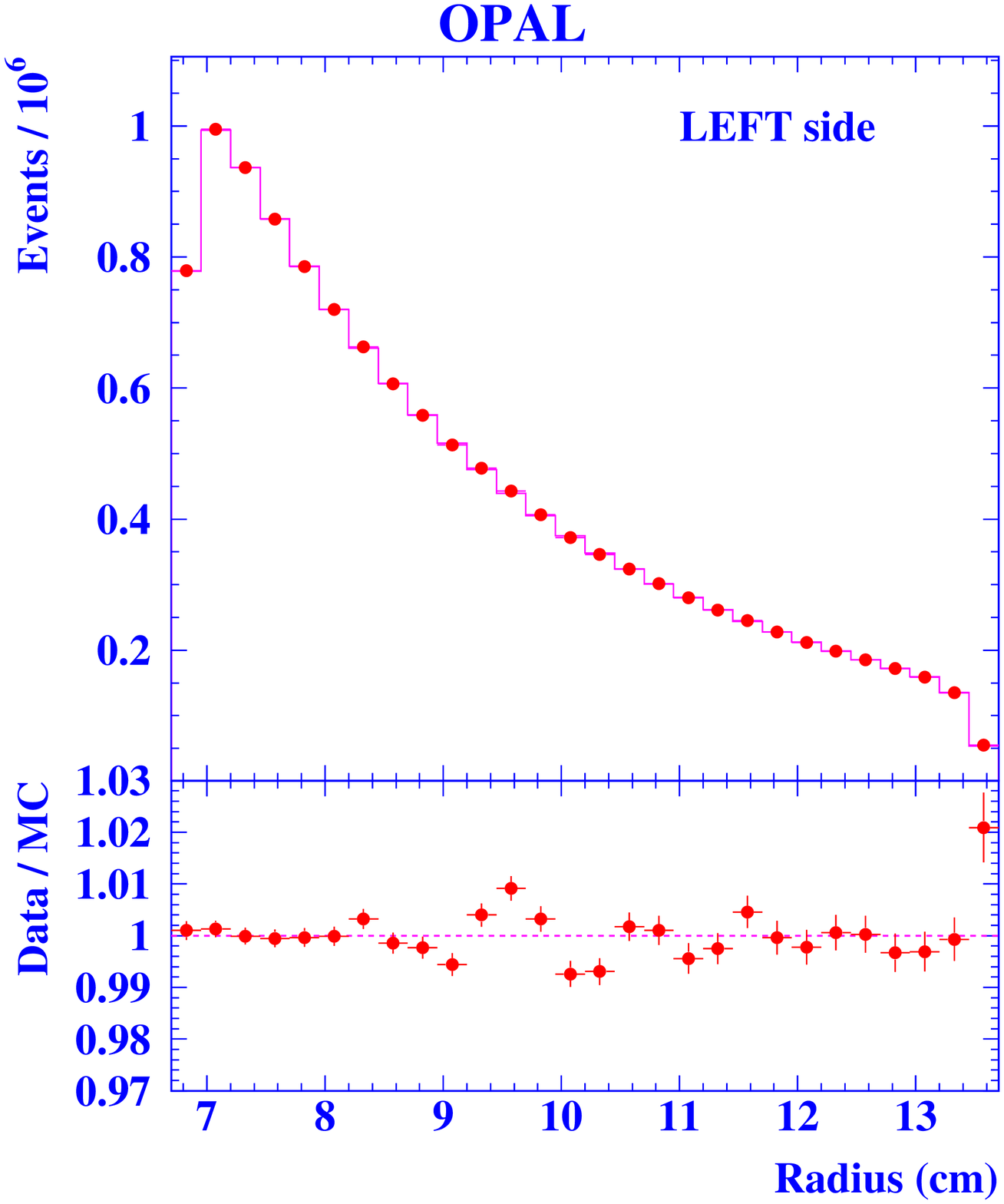}  &
\epsfxsize=0.46\textwidth\epsfbox[0 0 567 680]{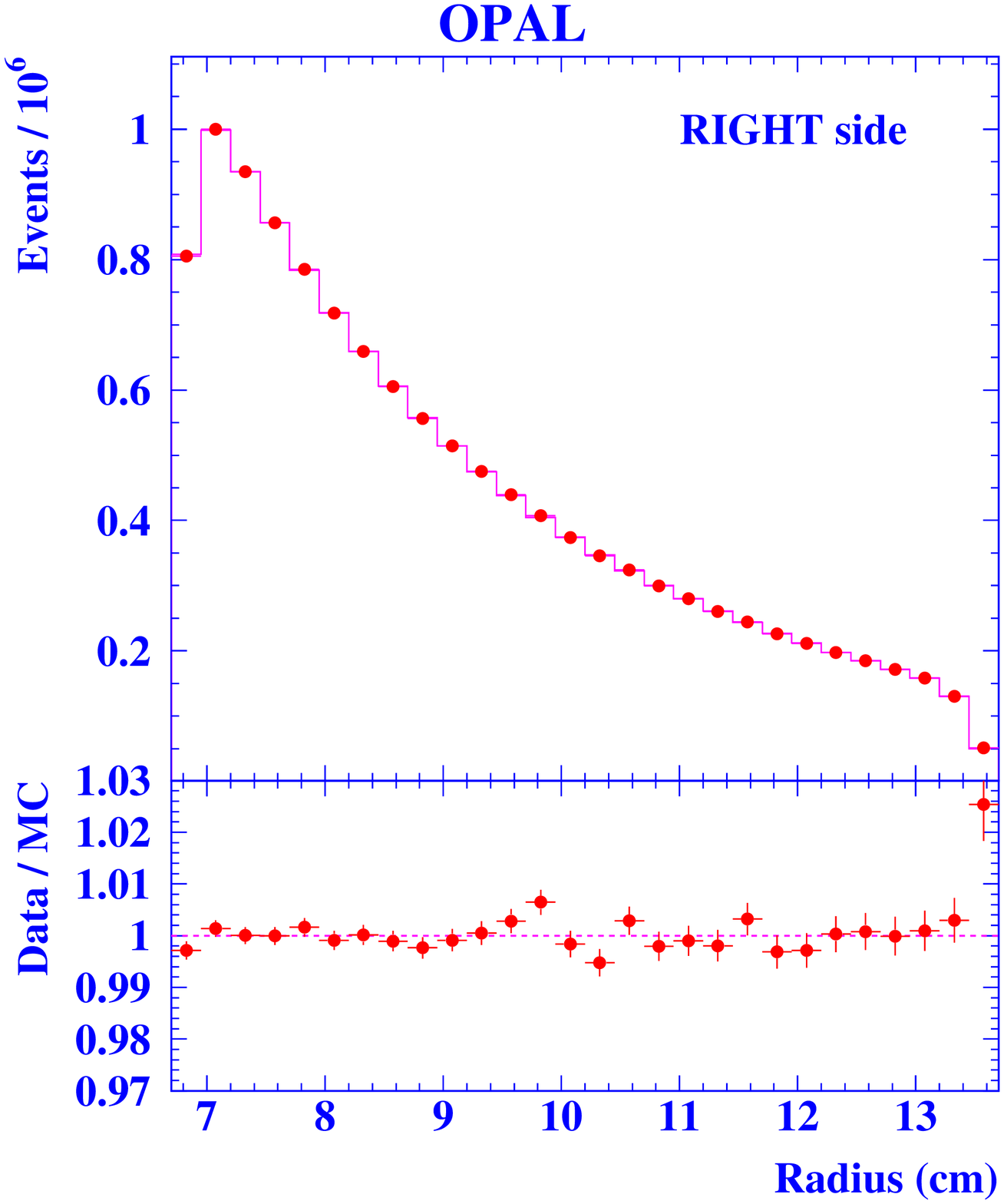}   \\
\end{tabular}
\end{center}
\caption
{Radial distributions for the complete data statistics. 
The points show the data and the histogram
the Monte Carlo prediction, assuming the expected running of $\alpha$,
normalized to the same number of events.
The lower plots show the ratio between data and Monte Carlo.}
\label{fig:dradial}
\end{figure}

The radial distributions are shown in Fig.~\ref{fig:dradial} 
for the complete data statistics, compared to the Monte
Carlo distributions normalized to the same number of events.
Due to the back-to-back nature of Bhabha events, the two sides do not contribute
independent statistical information. 
After the studies mentioned in section \ref{sec:sys},
we decided to use the Right side distribution for the final fits, 
to keep at minimum possible unassessed systematic errors.
Consistent results are obtained with the use of the Left side distribution.
%%%%%%%%%%%%%%%%%%%%%%%%%%%%%%%%%%%%%%%%%%%%%%%%%%%%%%%%%%%%%%%%%%%%%%%%%%%
%%%%%%%%%%%%%%%%%%%%%%%%%%%%%%%%%%%%%%%%%%%%%%%%%%%%%%%%%%%%%%%%%%%%%%%%%%%

\section{Fit method}
The counting rate of Bhabha events in the SiW is used to determine
the integrated luminosity, so that we cannot make an absolute measurement of 
$\alpha(t)$ without an independent determination of the luminosity.

We compare the radial distribution of the data (and hence the $t$-spectrum)
with the predictions of the BHLUMI Monte Carlo \cite{bhlumi}.
This is a multiphoton exponentiated generator accurate up to the 
leading logarithmic ${\cal O}(\alpha^2 L^2)$ 
terms \footnote{$L = \ln(|t|/m_\mathrm{e}^2)-1 \,$ is the large logarithm.}. 
Higher order
photonic contributions are partially included by virtue of the exponentiation.
It has been used to determine the luminosity at LEP and has been widely
cross-checked with many alternative calculations.
If the Monte Carlo is modified by setting the coupling to the constant value 
$\alpha(t) \equiv \alpha_0$, the ratio $f$
of the number of data to Monte Carlo events in a given radial bin is:
\begin{equation}
f(t) = \frac{N_{\mathrm{data}}(t)}{N_{\mathrm{MC}}^0(t)} \propto 
{ \left( \frac{1} {1-\Delta\alpha (t)} \right) }^2  \, .
\end{equation}
The dominant dependence of $\Delta\alpha (t)$ expected from theory is
logarithmic. 
We therefore fitted the ratio $f(t)$ as:
\begin{equation}
f(t) = a + b \, \ln \left( \frac{t}{t_0} \right)
\label{eq:Rfit}
\end{equation}
where $t_0 = -3.3$~GeV$^2$ is the mean value of $t$ in the data sample.
The parameter $a$, about unity, is not relevant 
since the Monte Carlo is normalized to the data.
The slope $b$ represents the full observable effect of the running of
$\alpha(t)$, both the leptonic and hadronic components. 
It is related to the variation of the coupling by:
\begin{equation}
\Delta\alpha(t_2) - \Delta\alpha(t_1) \simeq 
\frac{b}{2} \,\ln \left( \frac{t_2}{t_1} \right)
\label{eq:beff}
\end{equation}
where $t_1 = -1.81$~GeV$^2$ and $t_2 = -6.07$~GeV$^2$ 
correspond to the acceptance limits.

\section{Main systematic effects}
\label{sec:sys}
It is important to realize which systematic
effects could mimic the expected running or disturb the measurement.
The most potentially harmful effects are biases 
in the reconstructed radial coordinate. 
Most simply one could think of dividing the detector acceptance into two 
and determining the slope using only two bins.
In such a model the running is equivalent to a bias in the
central division of $70\,\mu$m. 
Biases on the inner or outer radial cut
have a little less importance and could mimic the full running for
$90$ or $210\,\mu$m systematic offsets respectively.
Concerning radial metrology, a uniform bias of $0.5\,$mm on all radii 
would give the same observable slope as the expected running.
Knowledge of the beam parameters, particularly the transverse offset and
the beam divergence, is also quite important.
Thus, limitation of systematic error in the reconstructed radial coordinate
is key to the current measurement.  

Details of how the coordinates are formed from the recorded pad information are 
found in \cite{lumipap}.
The fine radial and longitudinal granularity of the detector are exploited to
produce precise radial coordinates.
The reconstruction determines the radial coordinate of the highest energy
cluster, in each of the Right and Left calorimeters.
Each coordinate 
uses a large number of pads throughout the detector, from many
silicon layers, and is projected onto a reference layer,
close to the average longitudinal shower maximum.
The residual bias, or {\em anchor}, 
of this radial coordinate is then estimated at each pad boundary 
in a given layer of the detector. 
Here we rely on the fact that, on average, the pad with 
the maximum signal in any particular layer will contain the shower axis.
Then from the anchors we obtain bin-by-bin acceptance corrections 
which are applied to the radial distribution.
This procedure, named {\em anchoring}, 
is the most delicate part of the analysis, 
and was carefully studied \cite{alphat}. 
The challenging aspect is
controlling the residual bias on the radial coordinate to a level below
$\approx\!10\,\mu$m uniformly throughout the acceptance.

\section{Results}
The ratio of data to Monte Carlo is fitted 
to Eq.~\ref{eq:Rfit} and 
the results are given in Table~\ref{tab:fit.bello}. 
%%%%%%%%%%%%%%%%%%%%%%%%%%%%%%%%%%%%%%%%%%%%
\begin{table}[tb]
\caption[]
{Fit result for each dataset and average. 
For each value of $b$ the first error is
statistical and the second the full experimental systematic.
}
\begin{center}
\begin{tabular}{|l|c|r|c|}
\hline
Dataset & $\sqrt{s}$ & Number    & slope $b$\\
        &    (GeV)   & of events & $(\times 10^{-5})$ \\ 
\hline \hline
  93 $-2$   & 89.4510 &  879549 & $ 662 \pm 326 \pm \phz89$ \\
  93 pk     & 91.2228 &  894206 & $ 670 \pm 324 \pm \phz92$ \\
  93 $+2$   & 93.0362 &  852106 & $ 640 \pm 332 \pm \phz89$ \\
  94 a      & 91.2354 &  885606 & $ 559 \pm 326 \pm \phz86$ \\
  94 b      & 91.2170 & 4069876 & $ 936 \pm 152 \pm \phz71$ \\
  94 c      & 91.2436 &  288813 & $ \phz62  \pm 570 \pm 122$ \\
  95 $-2$   & 89.4416 &  890248 & $ 839 \pm 325 \pm 124$ \\
  95 pk     & 91.2860 &  581111 & $ 727 \pm 402 \pm 126$ \\
  95 $+2$   & 92.9720 &  885837 & $ 156 \pm 325 \pm 128$ \\
\hline \hline
  Average   & 91.2208 & 10227352 & $ 726 \pm \phz96 \pm \phz70$ \\
\hline
\multicolumn{3}{|l} {$\chi^2/$d.o.f. (stat.)}       &  $ 6.9/8$  \\
\multicolumn{3}{|l} {$\chi^2/$d.o.f. (stat.+syst.)} &  $ 6.5/8$   \\
\hline
\end{tabular}
\end{center}
\label{tab:fit.bello}
\end{table}
%%%%%%%%%%%%%%%%%%%%%%%%%%%%%%%%%%%%%%%%%%%%%%%%%%%%%%%%%%%%%%%%%%%%%%%
The nine datasets give consistent results, 
with $\chi^2$/d.o.f~$ = 6.9 / 8$ for the average $b$ considering only
statistical errors.
The most important systematic errors come from the anchoring procedure 
and the preshowering material, both affecting the radial coordinate.
The fit results are then combined, 
by considering the full error correlation matrix, obtaining:
\begin{displaymath}
b = (726 \pm 96 \pm 70 \pm 50) \times 10^{-5}
\end{displaymath}
where here, and also in the results quoted below,
the first error is statistical, the second is the experimental systematic and 
the third is the theoretical uncertainty.
The total significance of the measurement is $5.6 \,\sigma$. 

The theoretical uncertainty is dominated by the photonic corrections to the
leading $t$-channel diagram, in particular by missing 
${\cal O}(\alpha^2 L)$ terms, and the technical precision of the calculation. 
We estimated these uncertainties by
comparing BHLUMI with alternative Monte Carlo calculations.
Other uncertainties, from $\Z$ interference and the contribution of
light $\epem$ pairs, were also estimated and added in quadrature. 

%%%%%%%%%%%%%%%%%%%%%%%%%%%%%%%%%%%%%%%%%%%%%%%%%%%%%%%%%%%%%%%%%%%%%%%%%%%
\begin{figure}[t]
\begin{center}
\epsfxsize=0.7\textwidth
\epsfbox{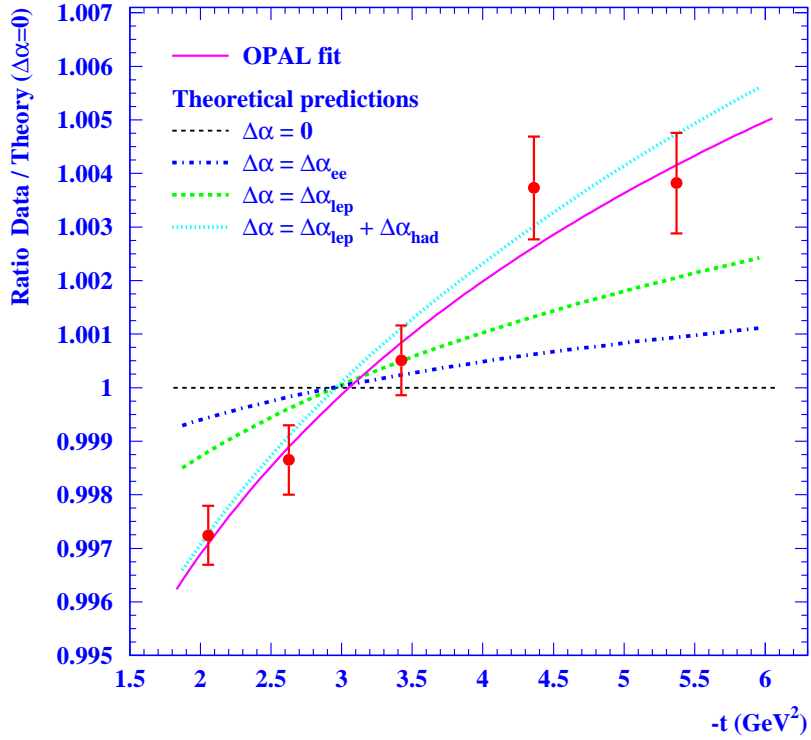}
\caption
{$|t|$ spectrum normalized to the BHLUMI theoretical prediction
for a fixed coupling ($\Delta\alpha = 0$).
The points show the combined OPAL data with statistical error bars.
The solid line is our fit. 
The horizontal line (Ratio=1) is the prediction if $\alpha$ were fixed. 
The dot-dashed curve is the prediction of running $\alpha$ 
determined by vacuum polarization with only virtual $\epem$ pairs, 
the dashed curve includes all charged lepton pairs and 
the dotted curve the full Standard Model
prediction, with both lepton and quark pairs.}
\label{fig:bella}
\end{center}
\end{figure}
The result for the combined data sample is illustrated in Fig.~\ref{fig:bella}.
The logarithmic fit to Eq.~\ref{eq:Rfit} describes the data very well,
$\chi^2$/d.o.f~$ = 1.9 / 3$,
although a simple linear fit would also be adequate, giving
$\chi^2$/d.o.f~$ = 2.7/ 3$.
The data are clearly incompatible with the hypothesis of a fixed coupling. 
The fitted logarithmic dependence agrees well 
with the full Standard Model prediction including both leptonic and
hadronic contributions, with the hadronic part obtained by the 
Burkhardt-Pietrzyk parameterization \cite{bp2001}.

The effective slope gives a measurement of the variation of the coupling 
$\alpha(t)$ from Eq.~\ref{eq:beff}:
\begin{displaymath}
\Delta\alpha(-6.07 \,\mathrm{GeV^2}) - \Delta\alpha(-1.81 \,\mathrm{GeV^2}) = 
(440 \pm 58 \pm 43 \pm 30) \times 10^{-5} \,.
\end{displaymath}
This is in good agreement with the Standard Model prediction, 
which gives $\delta\left(\Delta\alpha\right) = (460 \pm 16) \times 10^{-5}$
for the same $t$ interval, where the error originates from the uncertainty 
of the hadronic component. 
The evaluation \cite{bp2001} of $\Delta\alpha_{had}$
has a relative precision ranging from $2.5\,$\% 
at $t = -1.81 \,\mathrm{GeV^2}$ to 
$2.7\,$\% at $t = -6.07\,\mathrm{GeV^2} \,$ \cite{bolek}. 

The absolute value of $\Delta\alpha$ in our range of $t$
is expected to be dominated by $\epem$ pairs, with the relevant fermion species
contributing in the approximate proportions:
$\mathrm{e} : \mu : \mathrm{hadron} \simeq 4 : 1 : 2$.
Our measurement is sensitive, however, not to the absolute value of
$\Delta\alpha$, but only to its slope within our $t$ range.
Contributions to the slope $b$ in this range are predicted to be
in the proportion: $\mathrm{e} : \mu : \mathrm{hadron} \simeq 1 : 1 : 2.5$.
Fig.~\ref{fig:bella} shows these expectations graphically.
We can discard the hypothesis of running due only to virtual $\epem$ pairs with
a significance of $4.4\,\sigma$.

The data are also incompatible with the hypothesis of running due only to
leptons.
If we subtract the precisely calculable theoretical prediction for all
leptonic contributions,
$\delta(\Delta\alpha_{\mathrm{lep}}) = 202 \times 10^{-5}$, from the measured result, 
we can determine the hadronic contribution as:
\begin{displaymath}
  \Delta\alpha_{\mathrm{had}}(-6.07 \,\mathrm{GeV^2})
- \Delta\alpha_{\mathrm{had}}(-1.81 \,\mathrm{GeV^2}) = 
(237 \pm 58 \pm 43 \pm 30) \times 10^{-5} \,.
\end{displaymath}
This has a significance of $3.0\,\sigma$, considering all the errors.

\begin{figure}[htb]
\begin{center}
\epsfxsize=0.7\textwidth
\epsfbox{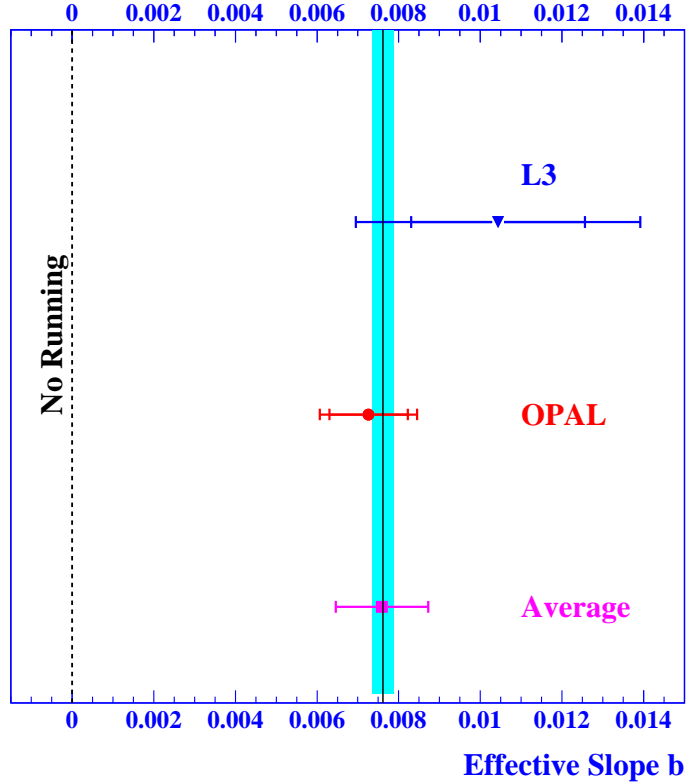}
\caption
{Effective slope $b = 2 <\mathrm{d}\Delta\alpha / \mathrm{d}\ln t>$.
The OPAL and L3 measurements are shown together with their average.
The solid line is the SM prediction, with the band showing its uncertainty. 
The dashed line at $b=0$ represents the case of no running.
}
\label{fig:opal-l3}
\end{center}
\end{figure}

Our result can be easily compared to the previous one by L3 \cite{l3}. If the
latter is expressed as a slope according to Eq.~\ref{eq:beff}, it becomes: 
$b^{(L3)} = (1044 \pm 348) \times 10^{-5}$.
The two measurements are shown in Fig.~\ref{fig:opal-l3}.
The L3 result has a larger error dominated by experimental systematics but is
consistent with ours. The average gives:
$b^{(ave)} = (759 \pm 113 \pm 50) \times 10^{-5}$, where the first error is
obtained from the experimental errors and the second is the theoretical
uncertainty that we estimated for our measurement, which will likely be 
common.
The average is in good agreement with the prediction 
using the Burkhardt-Pietrzyk parameterization.
%%%%%%%%%%%%%%%%%%%%%%%%%%%%%%%%%%%%%%%%%%%%%%%%%%%%%%%%%%%%%%%%%%%%%%%%%
%%%%%%%%%%%%%%%%%%%%%%%%%%%%%%%%%%%%%%%%%%%%%%%%%%%%%%%%%%%%%%%%%%%%%%%%%

\section{Conclusions}
We have measured the scale dependence of the effective QED coupling from the
angular distribution of small-angle Bhabha scattering at LEP, using the precise
OPAL Silicon-Tungsten luminometer.
We obtain the strongest direct evidence for the running of the QED coupling 
ever achieved in a single experiment, with a significance above $5\,\sigma$.
Moreover we report the first clear experimental evidence for the hadronic
contribution to the running in the space-like region,
with a significance of $3\,\sigma$.
This measurement is one of only a very few experimental tests of the running of
$\alpha(t)$ in the space-like region, where $\Delta\alpha$ has a smooth
behaviour. 
Our result is in good agreement with standard evaluations of $\alpha(t)$,
based on data in the time-like region.

\end{document}